\documentclass[twocolumn]{aastex631}
\usepackage{xspace}
\usepackage{amsmath}
\usepackage{soul}
\newcommand{\swift}{{\it Swift}\xspace}
\newcommand{\nicer}{\textit{NICER}\xspace}
\newcommand{\xmm}{{\it XMM-Newton}\xspace}
\newcommand{\chandra}{{\it Chandra}\xspace}
\newcommand{\cgs}{{\rm erg s$^{-1}$ cm$^{-2}$}\xspace}

\newcommand{\target}{{eRO-QPE1}\xspace}
\newcommand{\rx}{{RX~J1301.9+2747}\xspace}
\newcommand{\host}{2MASS~02314715-1020112}

\submitjournal{ApJ Letters}
\begin{document}

\title{Alive but Barely Kicking: News from 3+ years of \swift and \xmm X-ray Monitoring of Quasi-Periodic Eruptions from eRO-QPE1}
\correspondingauthor{Dheeraj R. Pasham}
\email{drreddy@mit.edu}
\author[0000-0003-1386-7861]{D.R. Pasham}
\affiliation{Kavli Institute for Astrophysics and Space Research, Massachusetts Institute of Technology, Cambridge, MA, USA}
\author[0000-0003-3765-6401]{E.R. Coughlin}
\affiliation{Department of Physics, Syracuse University, Syracuse, NY 13210, USA}
\author[0000-0001-6450-1187]{M. Zaja\v{c}ek}
\affiliation{Department of Theoretical Physics and Astrophysics, Faculty of Science, Masaryk University, Kotl\'a\v{r}sk\'a 2, 611 37 Brno, Czech Republic}
\author{Itai Linial}
\affiliation{Institute for Advanced Study, Einstein Drive, Princeton, NJ 08540, USA}
\author[0000-0002-4779-5635]{Petra Sukov\'a}
\affiliation{Astronomical Institute of the Czech Academy of Sciences, Bo\v{c}n\'{\i} II 1401, 141 00 Prague, Czech Republic}
\author[0000-0002-2137-4146]{C.J. Nixon}
\affiliation{School of Physics and Astronomy, Sir William Henry Bragg Building, Woodhouse Ln., University of Leeds, Leeds LS2 9JT, UK}
\author{Agnieszka Janiuk}
\affiliation{Center for Theoretical Physics, Polish Academy of Sciences, Al. Lotników 32/46, 02-668 Warsaw, Poland}
\author[0000-0003-2656-6726]{M. Sniegowska}
\affiliation{ School of Physics and Astronomy, Tel Aviv University, Tel Aviv 69978, Israel}
\affiliation{Nicolaus Copernicus Astronomical Center, Polish Academy of Sciences, ul. Bartycka 18, 00-716 Warsaw, Poland}
\author[0000-0002-9209-5355]{Vojt{\v e}ch Witzany}
\affiliation{Institute of Theoretical Physics, Faculty of Mathematics and Physics, Charles University, V Hole{\v s}ovi{\v c}k{\'a}ch 2, 180 00 Prague 8, Czech Republic}
\author[0000-0002-5760-0459]{V. Karas}
\affiliation{Astronomical Institute of the Czech Academy of Sciences, Bo\v{c}n\'{\i} II 1401, 141 00 Prague, Czech Republic}
\author{M. Krumpe}
\affiliation{Leibniz-Institut f\"ur Astrophysik Potsdam, An der Sternwarte 16, 14482 Potsdam, Germany}
\author[0000-0002-3422-0074]{D. Altamirano}
\affiliation{School of Physics and Astronomy, University of Southampton, UK}
\author[0000-0002-4043-9400]{T. Wevers}
\affiliation{Space Telescope Science Institute, 3700 San Martin Drive, Baltimore, MD 21218, USA}
\affiliation{European Southern Observatory, Alonso de Córdova 3107, Vitacura, Santiago, Chile}
\author{Riccardo Arcodia}
\affiliation{Kavli Institute for Astrophysics and Space Research, Massachusetts Institute of Technology, Cambridge, MA, USA}

\begin{abstract}
Quasi-periodic Eruptions (QPEs) represent a novel class of extragalactic X-ray transients that are known to repeat at roughly regular intervals of a few hours to days. Their underlying physical mechanism is a topic of heated debate, with most models proposing that they originate either from instabilities within the inner accretion flow or from orbiting objects. At present, our knowledge of how QPEs evolve over an extended timescale of multiple years is limited, except for the unique QPE source GSN 069. In this study, we present results from strategically designed \swift observing programs spanning the past three years, aimed at tracking eruptions from eRO-QPE1. Our main results are: 1) the recurrence time of eruptions can vary between 0.6 and 1.2 days, 2) there is no detectable secular trend in evolution of the recurrence times, 3) consistent with prior studies, their eruption profiles can have complex shapes, and 4) the peak flux of the eruptions has been declining over the past 3 years with the eruptions barely detected in the most recent \swift dataset taken in June of 2023. This trend of weakening eruptions has been reported recently in GSN~069. \edit1{However, because the background luminosity of eRO-QPE1 is below our detection limit, we cannot verify if the weakening is correlated with the background luminosity (as is claimed to be the case for GSN 069).} We discuss these findings within the context of various  proposed QPE models. 
\end{abstract}
\keywords{tidal disruption events, black holes, accretion disks}

\section{Introduction} \label{sec:intro}
Quasi-Periodic Eruptions (QPEs) are intense, repeating bursts of soft X-rays originating from nuclei of nearby galaxies \citep{gsn069, rxqpes, arcodia21}. Their central black holes have masses in the range of 10$^{5-7}$ M$_{\odot}$ as derived from host-galaxy stellar velocity dispersion scaling relations \citep{2022A&A...659L...2W}. There are currently four QPE systems known with recurrence periods (i.e., the time between subsequent flares) varying from a few hours to 0.8 days \citep{gsn069, arcodia21}. Two additional nuclear transients with one and a half \citep{2021ApJ...921L..40C} and a half eruption \citep{2023A&A...675A.152Q} have been suggested as potential QPE systems. While the first two (GSN~069 and \rx) were discovered in archival \xmm~ datasets, two systems (eRO-QPE1 and eRO-QPE2 as named in \citealt{arcodia21}) were found through follow-up of candidates from a systematic search in sky survey data from the {\it eROSITA} instrument onboard the Spectrum-Roentgen-Gamma (SRG) space observatory \citep{erosita}. These latter findings provide the exciting prospect of identifying more QPE sources with future all-sky X-ray surveys.

\begin{figure*}
    \centering
    \includegraphics[width=0.95\textwidth]{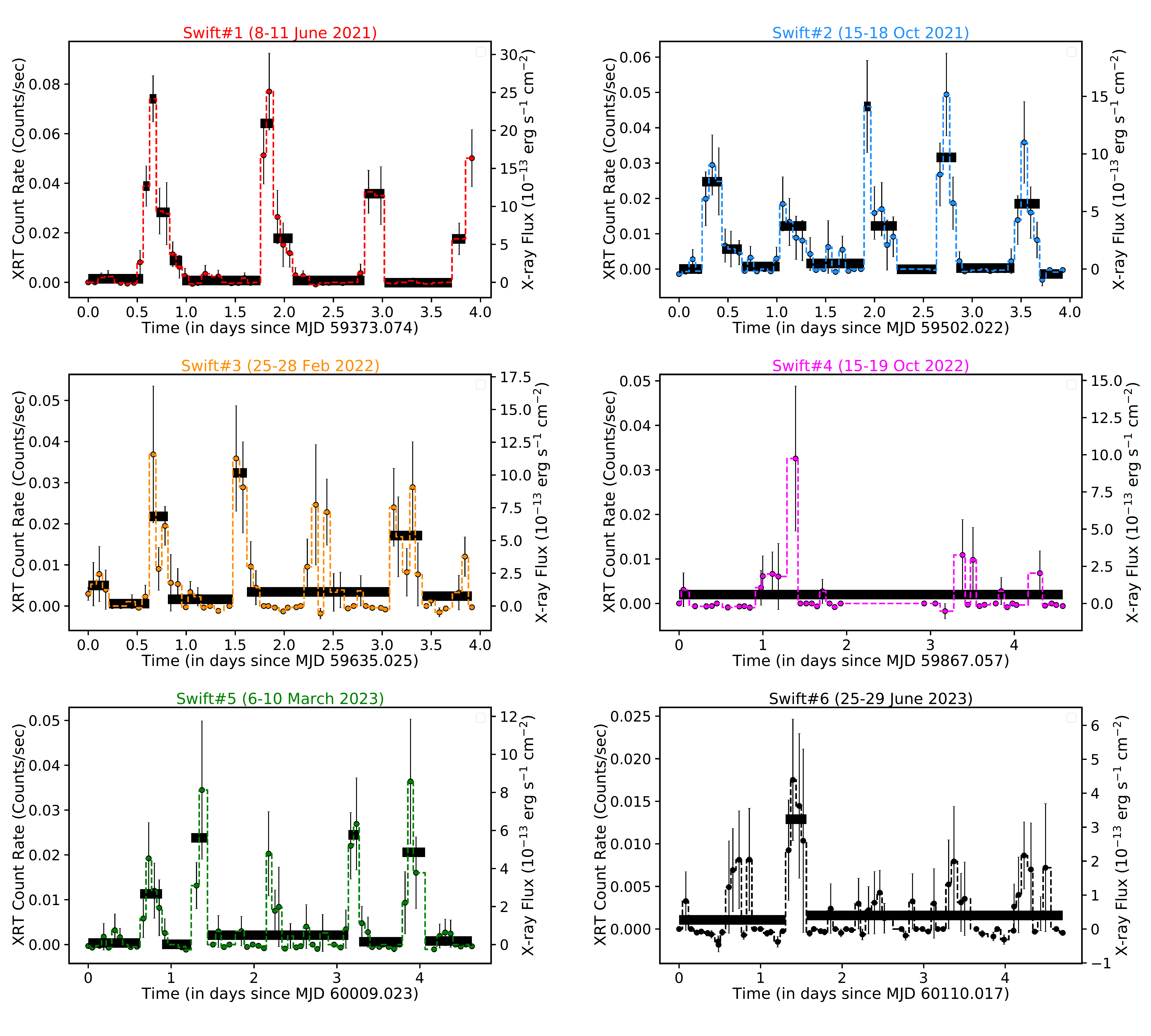}
    \caption{{\bf 0.3-1.2 keV X-ray light curves of \target.} Each light curve is from a high-cadence monitoring program with \swift. The observation dates are indicated at the top of each panel. Both the X-ray count rate and the observed flux were measured in the 0.3-1.2 keV band. The decrease in strength of eruptions over time is evident. Note that the y-scale is different in each panel. The thick black horizontal lines are the optimal time bins derived from the Bayesian blocks algorithm of \citealt{blocks2}.}
    \label{fig:lc}
\end{figure*}

Broadly speaking, QPEs have the following observational properties. They all have soft/thermal spectra with best-fit blackbody temperatures of a few 10s to a few 100s of eV \citep{arcodia21, gsn069}. There is no commonality in their burst profiles. For example, GSN~069, \rx and eRO-QPE2 have more or less symmetric bursts while eRO-QPE1 has a complex behavior where some eruptions show a fast-rise, slow decay behavior while others are more or less symmetric and can sometimes be broad \citep{2022A&A...662A..49A}. In general, their X-ray temperature is correlated with luminosity \citep{arcodia21, gsn069, rxqpes}. The recurrence time is not strictly periodic with GSN~069, \rx, \target and eRO-QPE2 showing variations on the order of $\approx$30\% \citep{miniutti23a, rxqpes}.

The underlying physical mechanism producing QPEs is currently unknown but an increased interest in this subject has led to several theoretical models being proposed in the last few years. These models can be put into two broad categories: ones that invoke inner accretion disk instabilities \citep{marzenamodel, kaurmodel, panmodel, Raj:2021b} perhaps similar to those occurring in stellar-mass black hole X-ray binaries GRS 1915+105 \citep{grsheartbeat} and IGR~J17091-3624 \citep{igr}, and those involving one or more orbiting stellar objects \citep{krolikmodel,itaimodel1, kingqpemodel, repeated_emris_model, itaimodel2, petra, alessiamodel, 2022A&A...661A..55Z, xianmodel}. It has also been proposed that self-lensing binary supermassive black holes (SMBHs) can, in principle, produce quasi-periodic flares, but this scenario appears to be inconsistent with data from GSN~069 and \rx~\citep{ingrammodel}. As evident in the  extensive list of theoretical models referenced above, the models with an orbiting object around a massive black hole have increased the excitement in the field as QPEs could potentially represent extreme mass ratio inspirals (EMRIs) of a secondary orbiter gradually sinking down to the central SMBH. If that is the case, some studies have suggested that some QPE sources could be detectable by future space-based gravitational wave missions like LISA and Tianqin \citep{2022A&A...661A..55Z}. But \cite{qpelisanodet} have argued that the signals from the currently known QPE sources may be weak to be detectable by future gravitational wave detectors. Irrespective of the underlying mechanism, QPEs have opened up a unique new window into the inner accretion flows of massive black holes. 

To pin down the mechanism driving QPEs, observations constraining their long-term evolution are necessary. Such information is available in the published literature only for GSN~069 \citep{miniutti23a, alivenkicking}. Here we present results from a monitoring campaign over an extended temporal baseline of three years using {\it Neil Gehrels Swift} (\swift hereafter) and archival \xmm observations of eRO-QPE1/\host$\footnote{We follow the naming convention of  \citealt{arcodia21}}$ which exhibited QPEs separated by $\approx$0.8 d during the first monitoring dataset \citep{arcodia21}. We performed additional high-cadence observations on multiple epochs, and our main observational findings are discussed in section \ref{sec:results}. \edit1{We discuss the implications of our findings within the context of several proposed theoretical models in section \ref{sec:models} and compare with GSN~069 and discuss future prospects of tracking \target in section \ref{sec:discussion}. We summarize our findings in section \ref{sec:summary}.}

\begin{table*}
\centering
\begin{tabular}{ccccccccc}
{\bf Epoch} & {\bf MJD$_{\rm start}$} & {\bf MJD$_{\rm end}$} & {\bf Count Rate} & {\bf Counts} & {\bf Average kT} & {\bf Average}                 & {\bf Peak}                    & {\bf Quiescence} \\
             & (days)                     & (days)                   &     (counts/sec)  &               & (keV)              & Flux & Flux & Flux \\\hline
Swift\#1 & 59373.074 & 59376.986 & 0.0244$\pm$0.0016 & 223& 0.13$\pm$0.01& 8.0$\pm$0.7& 20.6$\pm$3.1& $<$0.3 \\
Swift\#2 & 59502.022 & 59505.948 & 0.0126$\pm$0.0012 & 124& 0.13$\pm$0.01& 3.9$\pm$0.4& 11.5$\pm$2.2& $<$0.4 \\
Swift\#3 & 59635.025 & 59638.941 & 0.0099$\pm$0.0011 & 94& 0.097$\pm$0.012& 3.1$\pm$0.6& 9.7$\pm$2.8& $<$0.3 \\
Swift\#4 & 59867.057 & 59871.634 & 0.0058$\pm$0.0016 & 15& 0.097$\pm$0.012& 1.7$\pm$0.6& 6.5$\pm$3.9& $<$0.4 \\
Swift\#5 & 60009.023 & 60013.651 & 0.0096$\pm$0.0012 & 64& 0.12$\pm$0.016& 2.3$\pm$1.1& 7.1$\pm$3.8& $<$0.2 \\
Swift\#6 & 60110.017 & 60114.723 & 0.0043$\pm$0.0008 & 33& 0.12$\pm$0.016& 1.1$\pm$0.4& 3.2$\pm$1.7& $<$0.3 \\
\hline
\end{tabular}
{\caption{{\bf Summary of \target's \swift/XRT data and spectral modeling of its eruptions}. Here 0.3-1.2 keV \swift/XRT spectra were fit with {\it tbabs*ztbabs*zashift(diskbb)} model using {\it XSPEC} \citep{xspec}. {\bf MJD$_{\rm start}$} and {\bf MJD$_{\rm end}$} represent the start and the end times (in units of MJD days) of the \swift monitoring campaign. {\bf Count Rate} and {\bf Counts} represent background-subtracted values in 0.3-1.2 keV. The column density at the host, {\it ztbabs}, was fixed at the best-fit \xmm value of 0.069$\times$10$^{22}$ cm$^{-2}$. Temperatures of \swift\#1 and \swift\#2, \swift\#3 and \swift\#4, and \swift\#5 and \swift\#6 were tied. All the errorbars represent 1-$\sigma$ uncertainties except for the 3$\sigma$ quiescence level upper limits. The {\bf Average Flux}, {\bf Peak Flux} and {\bf Quiescence Flux} values correspond to 0.3-1.2 keV and have units of 10$^{-13}$ erg s $^{-1}$ cm$^{-2}$. The total C-stat/degrees of freedom was 31.6/34. }\label{tab: tab1}}
\label{tab:tab1}
\end{table*}

\section{Data and Observational Findings}
\label{sec:results}
\swift's X-Ray Telescope (XRT; \citealt{xrt}) performed six sets (\swift\#1...6; Table \ref{tab:tab1}) of high-cadence monitoring observations of \target. A detailed discussion of data reduction and spectral analyses is discussed in Appendix section \ref{sec:data}. Here we highlight the main observational findings.

\begin{figure}
    \centering
    \hspace{-0.1in}
    \includegraphics[width=0.45\textwidth]{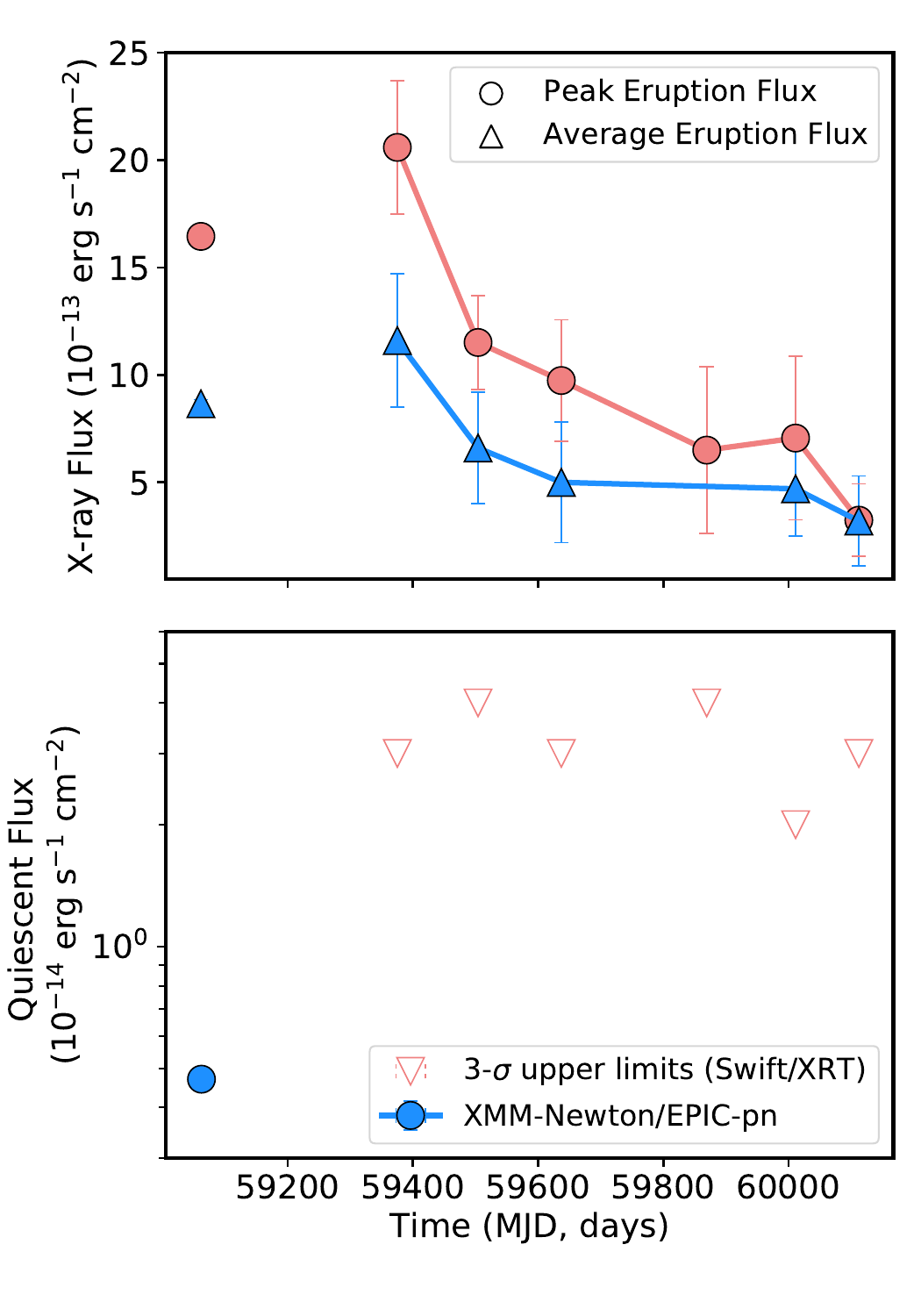}
    \caption{{\bf Long-term Evolution of the average peak luminosity of the eruptions (top) and the quiescent level (bottom).} The fluxes are observed values in the 0.3-1.2 keV band. The errorbars represent 1$\sigma$ uncertainties. They include both measurement and model fitting uncertainties (see section \ref{sec:xrt}). \swift/XRT data were not sensitive enough to detect the quiescent level but 3$\sigma$ upper limits are shown. \xmm's observed 0.3-1.2 keV quiescent flux was derived from combining data from obsIDs 0861910201 and 0861910301 (blue point in the bottom panel). }
    \label{Fig:fig3}
\end{figure}

Eruptions were detected in all these campaigns but with decreasing strength over time. This is evident even by eye in Fig.~\ref{fig:lc} where the brightness at peaks is gradually decreasing with time. The same is quantified in the top panel of Fig.~\ref{Fig:fig3} \textbf{which shows that the peak and average fluxes of the eruptions have decreased by approximately a factor of 10 and 4, respectively, over three years. } The quiescent level was detected in the first \xmm observation but only upper limits were available during the \swift observations (Fig.~\ref{Fig:fig3} bottom panel). 

\begin{figure*}
    \centering
    \includegraphics[width=0.99\textwidth]{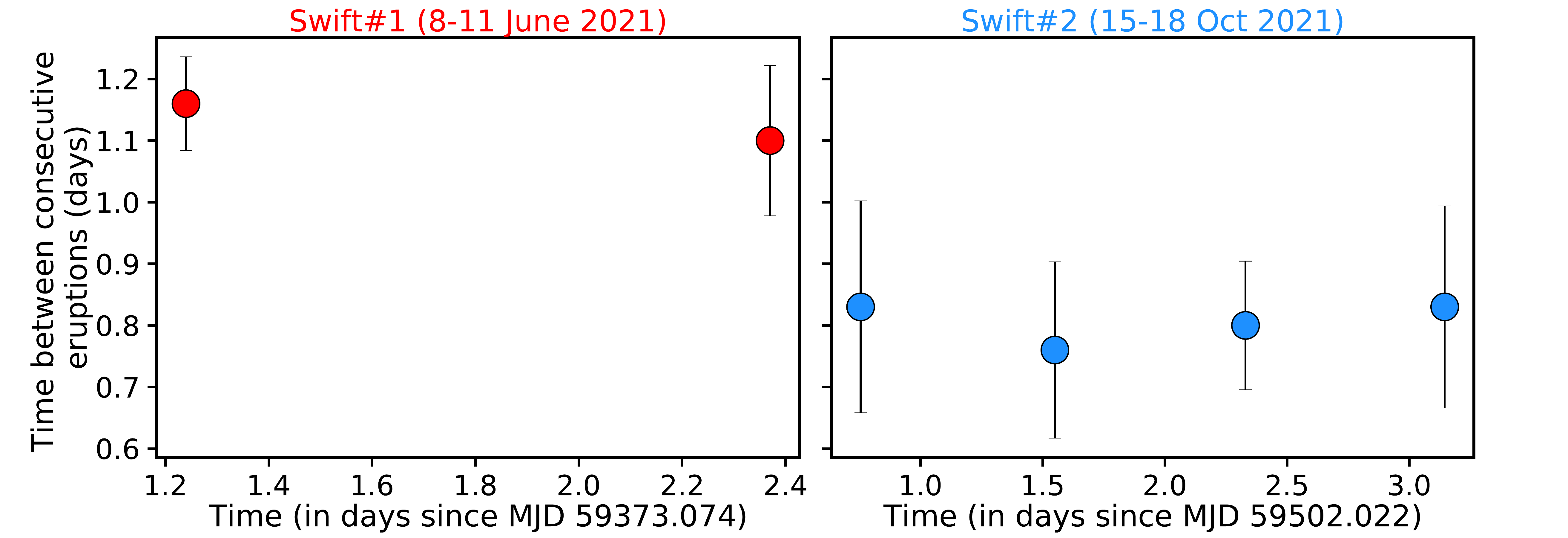}
    \caption{{\bf Evolution of time between consecutive eruptions with time for \swift\#1\&2 datasets.} The peaks of eruptions were determined in a model-independent manner using the Bayesian blocks algorithm of \citealt{blocks2}. These values are consistent with the Lomb Scargle peaks shown in Fig. \ref{fig:lsps}. }
    \label{fig:delT}
\end{figure*}

\begin{figure}
    \centering
    \includegraphics[width=0.43\textwidth]{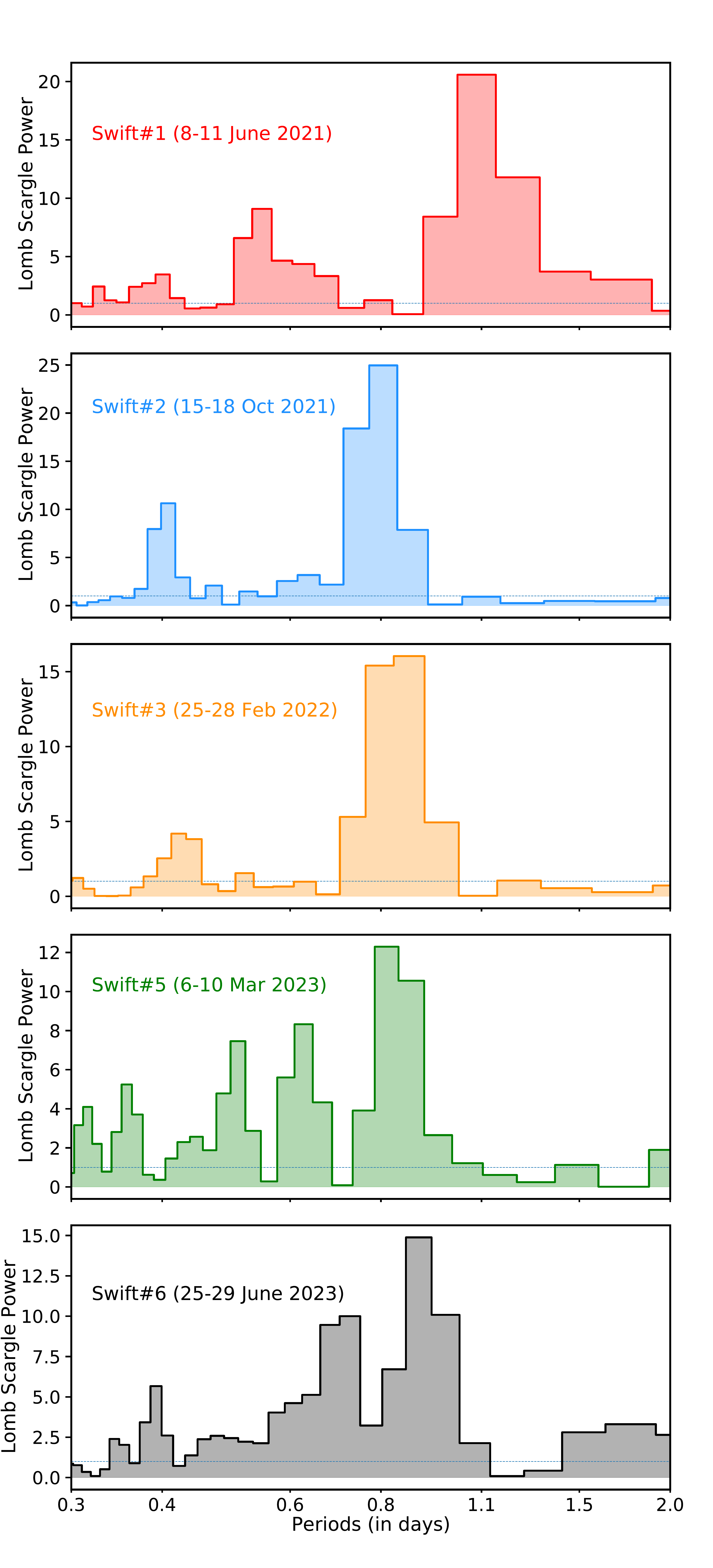}
    \caption{{\bf Lomb Scargle Periodograms of the light curves shown in Fig.~\ref{fig:lc}.} Here were zoom in on the 0.3-2.0 days timescale. The highest peaks (corresponding full width at half maximum) for \swift\#1, \swift\#2, \swift\#3, \swift\#5, and \swift\#6 are 1.09$^{+0.23}_{-0.08}$~d, 0.81$^{+0.03}_{-0.10}$~d, 0.83$^{+0.09}_{-0.07}$~d, 0.82$^{+0.10}_{-0.04}$~d, and 0.91$^{+0.12}_{-0.25}$~d, respectively. We do not show the LSP from \swift\#4 as it was affected by a large data gap. The blue/dashed horizontal line at a power value of 1 represents the LSP's nominal white noise level. While the main purpose of these plots is to show the evolution of the main LSP peak and not to delve into the details of the LSP, it is interesting that a harmonic is also present in all cases at roughly 1/2 the period of the main peak.}
    \label{fig:lsps}
\end{figure}

Another key observational finding is that the recurrence time varies from one monitoring campaign to the next. Surprisingly, the three eruptions seen in the first \swift campaign (\swift\#1) were separated by $\sim$1.1 days (left panel of Fig. \ref{fig:delT}), i.e., $\sim$40\% longer than previously published recurrence time of 0.8 days \citep{arcodia21}. However, this recurrence time returned to a mean value of 0.8 d roughly six months later in Oct 2021 (\swift\#2; right panel of Fig. \ref{fig:delT}). This is also quantified in the Lomb Scargle Periodograms shown in  Fig.~\ref{fig:lsps}. Over the course of all the six monitoring campaigns, the recurrence time varied between 0.6 and 1.2 days (see section \ref{sec:xrt}  and Fig.~\ref{fig:lsps}).

While several eruptions reported here are consistent with a fast-rise and smooth decay profile reported in \citealt{arcodia21}, there are numerous examples of more complex profile shapes. For example, the second eruption in \swift\#2 has a extended decay. The 4$^{th}$ and the 5$^{th}$ eruptions (around days 2.3 and 3.3) in \swift\#3 have a dip near their peaks. The eruption around day 1.5 in \swift\#6 and the last two eruptions in \swift\#2 appear symmetric in shape. In summary, contrary to previous reports, we find that the eruption profiles have different shapes. This is an important aspect to stress
as \citet{arcodia21} used the fast-rise and slow decay profile as an argument against a certain type of radiation pressure instability.

\section{Implications for various theoretical models}\label{sec:models}
We now discuss the observations presented above within the context of various models proposed for QPEs. 

\subsection{QPEs models with orbiting objects}
\subsubsection{Repeating partial Tidal Disruption Event (TDE)}
First, we explore three flavors of repeating partial TDE (rpTDE) scenarios: 1) \target's central black hole is repeatedly disrupting a star on a timescale of several years and the QPEs are produced as a result of inner disk-related physics, 2) QPEs are a direct result of a white dwarf that is in a $\sim$1 day orbit around an SMBH, and 3) rpTDE of a main sequence star by an SMBH. 

{\citet{miniutti23a} noted that GSN~069's long-term behavior appears to be consistent with the tidal disruption of a star, and 
the QPEs only appeared once the flux fell below a critical value. It may therefore be the case that the QPE phenomenon is intricately tied to, and in fact requires, a prior TDE, with the production mechanism related to the ensuing disk physics (and any potential instabilities associated therewith) or a change in the morphology of the returning debris stream \citep{coughlin20, guolo23}. 
Unfortunately, the quiescent X-ray flux from eRO-QPE1 is below the \swift/XRT detectability threshold (see Figure \ref{Fig:fig3}), and we are not presently able to test this hypothesis. However, future \xmm monitoring observations can address the nature of the long-term evolution of quiescence emission (see section \ref{sec:future}). 

A second scenario has been suggested in which the X-ray eruptions are produced due to accretion following the repeated partial tidal stripping of a white dwarf by the black hole in the nucleus of the galaxy (\citealt{kingqpemodel}; see also \citealt{zalamea10} in the context of Extreme Mass Ratio Inspirals/EMRIs). If the energy generated due to accretion is $E_{\rm acc} = \eta {M}c^2$ with $\eta = 0.1$, then using a luminosity distance of 233 Mpc alongside the mean observed eruption flux of 3.4$\times$10$^{-13}$ \cgs (see Figure \ref{Fig:fig3}) implies an X-ray (0.3--1 keV) energy release per outburst of $E \simeq 7\times 10^{46}$ erg (average flux$\times$4$\times\pi$(luminosity distance)$^{2}\times(1+z=0.0505)\times$(average eruption duration of 8 hours from \citealt{arcodia21})), and hence an accreted mass of $M_{\rm acc} = E_{\rm acc}/(0.1 c^2) \simeq 4\times 10^{-7} M_{\odot}$. For a white dwarf with mass $\sim {\rm few}\times 0.1 M_{\odot}$, this amounts to a very small fraction of the total mass of the star, implying that the pericenter distance of the white dwarf is extremely fine-tuned to coincide with its partial tidal disruption radius (i.e., where material is just barely able to be removed from the surface of the star). How the star achieved precisely this distance is not clear, as neither gravitational-wave emission nor tidal interactions can dramatically change the pericenter distance. However, because the amount of mass stripped from the star is a very sensitive function of distance (see \citealt{guillochon13}), we might expect variation in the mass accreted -- and hence the luminosity (according to this model) -- even for small changes to the pericenter due to, e.g., the exchange of angular momentum between the star and the orbit. These secular changes are consistent with \target's trend in Figure \ref{Fig:fig3}.

Additionally, \target's black hole is estimated to be $\sim 10^{6} M_{\odot}$ \citep{wevers23}, which implies that the pericenter distance of the star must be highly relativistic -- at most on the order of a few gravitational radii. Adopting a white dwarf mass of $0.6M_{\odot}$ and a corresponding radius of $R_{\star} = 0.011R_{\odot}$ \citep{nauenberg72} yields a tidal disruption radius of $0.6 GM_{\rm BH}/c^2$, i.e., the partial tidal disruption radius is about twice that value \citep{guillochon13} at $\sim 1.2 GM/c^2$. If we set the pericenter distance of the star to $1.2 GM/c^2$ (which requires a rapidly spinning black hole and a prograde orbit for the star to not be directly captured), and the semi-major axis is determined from the recurrence period of $T = 1$ day to be $a = (T\sqrt{GM}/(2\pi))^{2/3} \simeq 200 GM/c^2$ for $M = 10^6M_{\odot}$, then this corresponds to an eccentricity of $e \simeq 0.994$. For these parameters, the gravitational-wave inspiral time can be estimated from Equations 5.6 and 5.7 from \citet{peters64}, and is $\sim 3.2$ years, i.e., the source would have declined substantially in recurrence time during the three years over which it has been observed. This is in obvious disagreement with the behavior exhibited by eRO-QPE1, which has a recurrence time that is relatively stable (see Figure \ref{fig:lsps}). 

More generally, we can calculate the time-dependent evolution of the orbital parameters (i.e., the semimajor axis and the eccentricity) by integrating Eqs.\ (5.6)--(5.7) of \citet{peters64} for different black hole and stellar masses. Figure \ref{fig:period_decay} shows the orbital period of the white dwarf as a function of time in years for four different black hole masses; the left panel uses a white dwarf mass of $M_{\star} = 0.6 M_{\odot}$ (also used by \citealt{kingqpemodel} and where the mass distribution of white dwarfs peaks), while the right panel adopts $M_{\star} = 0.4 M_{\odot}$. Here we assumed an initial orbital period of 1 day, a pericenter distance equal to twice the tidal radius $r_{\rm t} = R_{\star}\left(M_{\bullet}/M_{\star}\right)^{1/3}$, and the mass-radius relationship $R_{\star} = 0.011 \left(M_{\star}/(0.6 M_{\odot}\right)^{-1/3} R_{\odot}$, which is valid for a non-relativistic (i.e., for a lower-mass white dwarf) fully degenerate gas. We use the leading-order Newtonian estimates for the tides, gravitational radiation, and orbital dynamics. \textbf{Since the orbit is highly relativistic, there will be substantial corrections beyond the lowest-order solution given in \citet{peters64} (see, e.g., \citealt{Blanchetlrr,2021PhRvD.104j4023T}). These higher-order terms generally lead to accelerated decay of the eccentricity and period shortening, especially for the case of heavier central SMBH, even though the qualitative picture stays the same. Thus, the results shown in Figure \ref{fig:period_decay} should be considered upper limits to the orbital period as a function of time, i.e., higher-order terms will only result in an acceleration in the decay rate of the period.} 

\textbf{This figure shows} that for a white dwarf mass of $M_{\star} = 0.6 M_{\odot}$, the orbital period decays substantially -- by at least a factor of 2 -- \textbf{until the black hole mass is} well below the value inferred from the $M-\sigma$ relation \textbf{and into the intermediate-mass black hole regime}. For $M_{\star} = 0.4 M_{\odot}$ the period decays by at least $\sim 15\%$ \textbf{for $M_{\bullet} = 10^5 M_{\odot}$, and only for $M_{\bullet} = 10^4 M_{\odot}$ and $M_{\bullet} = 10^3 M_{\odot}$, i.e., intermediate-mass black holes, is the period change sufficiently small that it would remain undetected over the three-year observational period of \target}. This figure shows that, if \target's eruptions are produced from the repeated tidal stripping of a white dwarf, the black hole mass must be substantially smaller than the one inferred from $M-\sigma$, or the white dwarf mass must be very small (making the star rare). 

We also note that, for partial disruptions in which $\gtrsim 10\%$ of the mass of the star is removed, which corresponds to pericenter distances smaller than $\sim r_{\rm t}/(0.65)$ for a 5/3-polytropic star \citep{guillochon13, mainetti17, miles20}, higher-order moments (i.e., beyond the quadrupole) of the gravitational field of the black hole induce a positive-energy kick to the surviving star (and this has been verified across a wide range of black hole masses and stellar/planetary types; e.g., \citealt{faber05, manukian13, gafton15, kremer22}). This effect could conceivably stabilize the orbital decay that is induced by the gravitational-wave emission. However, for the extremely small amount of mass lost by the star necessary to power QPEs, the stellar pericenter distance is such that the tidal excitation of modes in the star removes energy from the orbit (i.e., the star is in the classic tidal dissipation regime; \citealt{fabian75, press77}). Specifically, it was recently shown by \citet{cufari23} that, for a pericenter distance of $r_{\rm t}/(0.55)$ -- very close to the distance from the black hole at which almost no mass is lost from the star -- the change in the specific energy of the orbit is $\Delta \epsilon \simeq -0.025 GM_{\star}/R_{\star}$ and is roughly the maximum amount by which tides reduce the energy of the orbit (i.e., larger pericenter distances yield less tidal excitation, but smaller pericenter distances lead to a positive-energy kick; see Figure 1 in \citealt{cufari23}). In this case, the fractional change in the energy of the orbit is $\Delta \epsilon/\epsilon = 0.025 M_{\star} a/(M_{\bullet}R_{\star}) \sim 0.1\%$ for $M_{\star} = 0.6 M_{\odot}$, $R_{\star} = 0.011 R_{\odot}$, $M_{\bullet} = 10^6M_{\odot}$, and $a = 200 GM_{\bullet}/c^2$, which is comparable in magnitude to the per-orbit change in energy induced by gravitational-wave emission. Thus, in this case we would only expect tides to accelerate the inspiral. 

\begin{figure*}
    \includegraphics[width=0.495\textwidth]{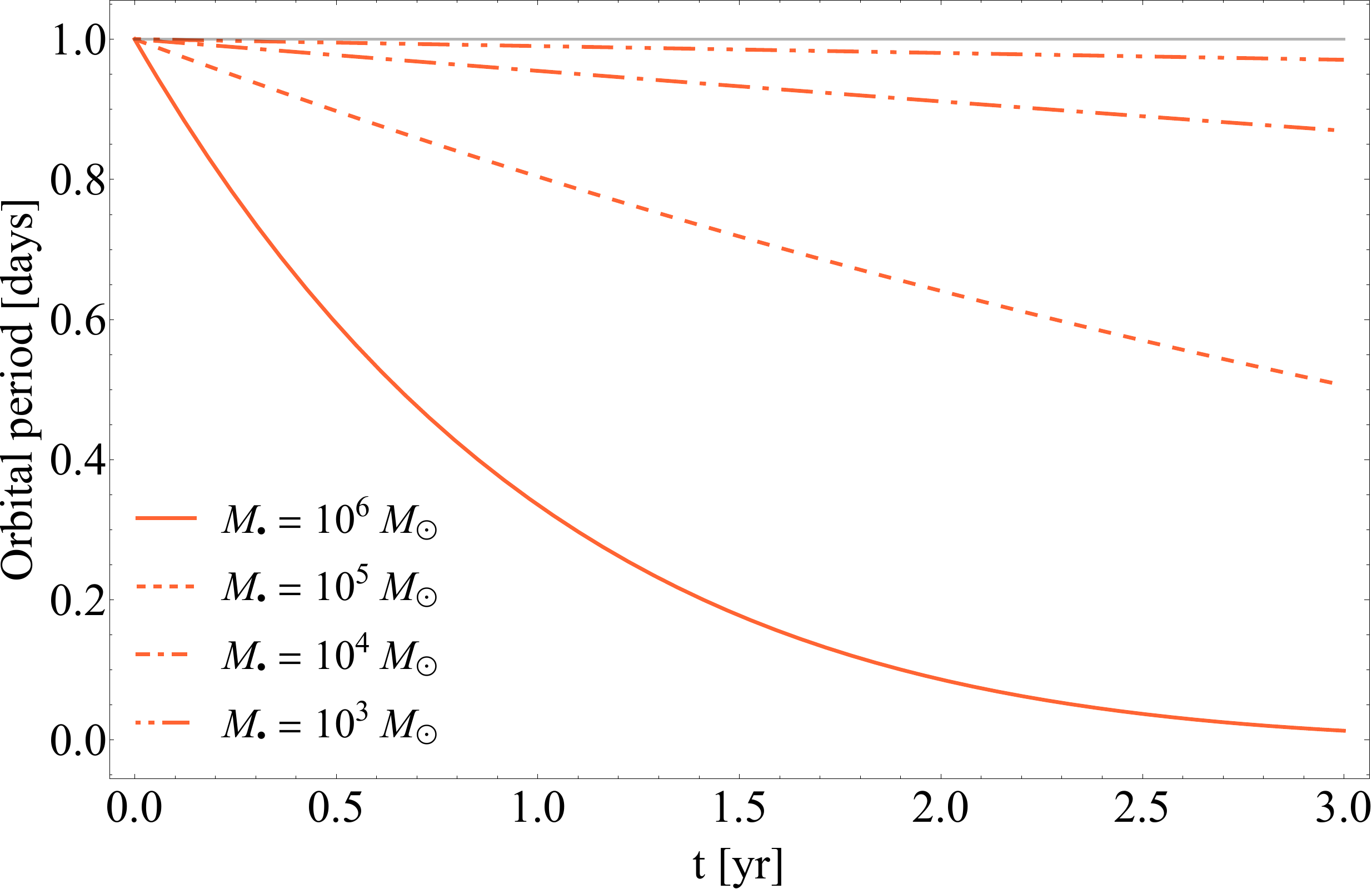}
    \includegraphics[width=0.495\textwidth]{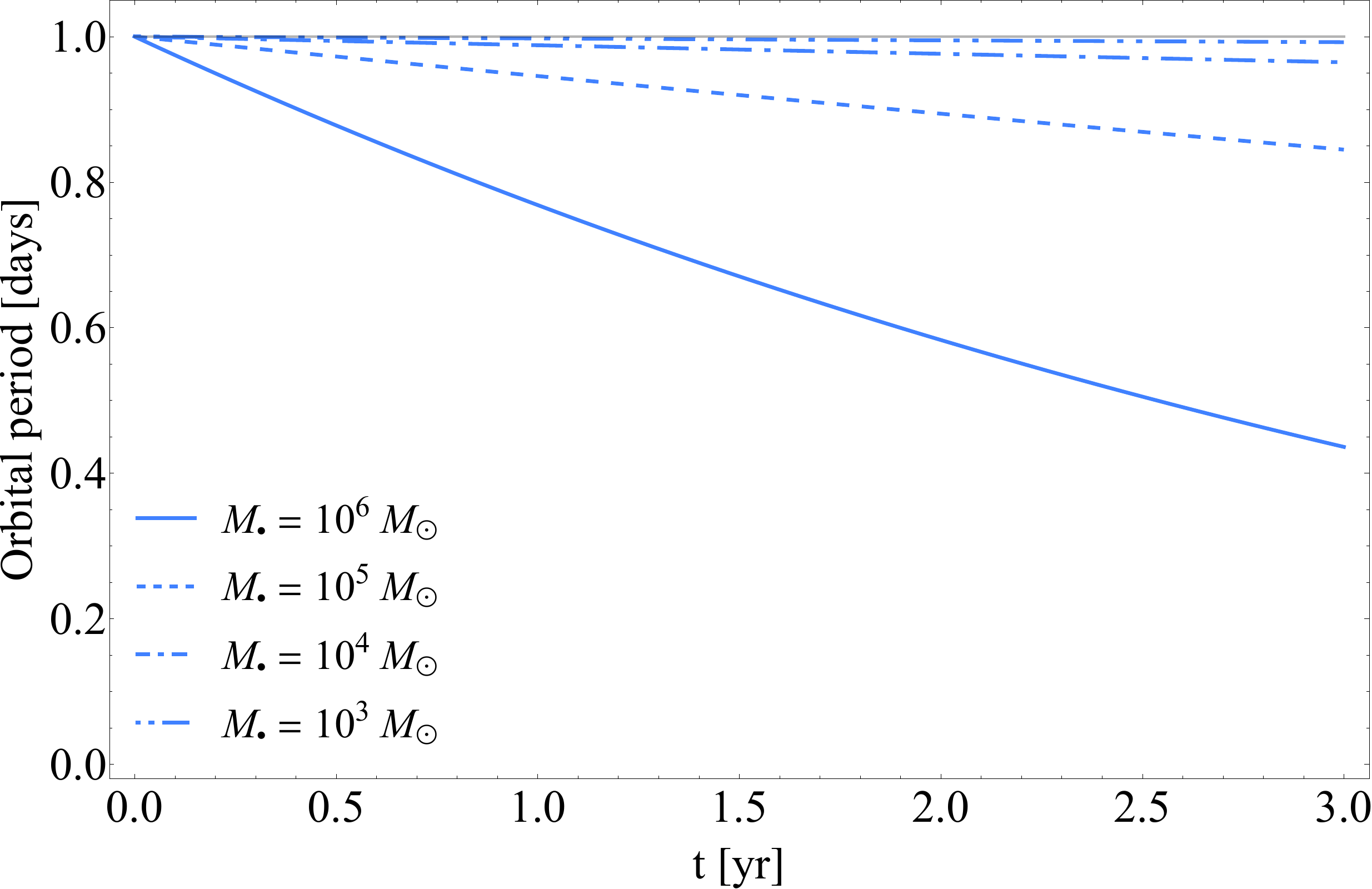}
    \caption{The Keplerian orbital period of a white dwarf orbiting a massive black hole, with black hole mass indicated in the legend, as a function of time in years, where the left (right) panel adopts a white dwarf mass of $M_{\star} = 0.6 M_{\odot}$ ($0.4 M_{\odot}$). The pericenter distance is equal to twice the canonical tidal radius, while the orbital period is initially equal to the recurrence time of 1 day. Over the course of the observations of \target, being roughly 3 years, there would be a substantial and noticeable decline in the recurrence time of the flares owing to gravitational-wave emission, unless the black hole mass is significantly below the value inferred from the $M-\sigma$ relation (being $\sim 10^6M_{\odot}$), or the white dwarf mass is significantly smaller than the mean value of $\sim 0.6 M_{\odot}$.}
    \label{fig:period_decay}
\end{figure*}

In the third scenario of repeated partial disruption of a main sequence star, it is also difficult to reconcile the observed timescales and energetics of QPEs. This model has been invoked to explain the repeating nuclear transients ASASSN-14ko \citep{payne21}, AT2018fyk \citep{wevers23}, eRASSt-J0456 \citep{liu23}\textbf{, and \emph{Swift} J0230 \citep{evans23, guolo23}}. Specifically, we would expect the return time of the tidally stripped debris to be comparable to the dynamical time at the surface of the star, multiplied by the square root of the mass ratio of the black hole to the star \citep{lacy82, rees88}. For a white dwarf, this timescale is of the order of a ksec -- in rough agreement with the flare duration of QPEs -- while for main sequence stars it is $\sim few\times 10$ days (see, e.g., the simulations in \citealt{guillochon13, golightly19, lawsmith20, nixon21}), which is orders of magnitude longer than the flare duration in QPEs, in general. Similar to the white dwarf model, the extremely small amount of mass accreted per burst is also problematic, and requires fine tuning to achieve a pericenter distance that is very closely aligned with the partial disruption radius of a star.


\subsubsection{QPEs from interactions of an orbiting perturber with the accretion disk}
\citet{petra} proposed that QPEs could be produced from repeated interaction of an object with the accretion disk of an SMBH \citep[see also][]{xianmodel,alessiamodel,itaimodel2}. In their model, at each interaction, the perturbation causes: 1) the modulation of the accretion rate onto the black hole depending the ratio of the influence radius of the object to its distance, and 2) ejection of matter clumps towards the magnetic poles, which can drive a quasiperiodic ultrafast outflow. Assuming that the perturber period is equal to twice the eruption period (two eruptions per orbit, i.e. $P_{\rm orb}\sim 2\times 0.9$ days), the semi-major axis is given by $a\sim 399.3 (P_{\rm orb}/1.8\,\text{days})^{2/3}(M_{\bullet}/10^{5.8}\,M_{\odot})^{-2/3} R_{\rm g}$, where $R_{\rm g}=GM_{\bullet}/c^2$ is a gravitational radius of the SMBH. Hence, it would vary between $\sim 1003\,R_{\rm g}$ for $M_{\bullet}=10^{5.2}\,M_{\odot}$ and $\sim 159\,R_{\rm g}$ for $M_{\bullet}=10^{6.4}\,M_{\odot}$.

The varying recurrence time in Fig. \ref{fig:lsps} and \ref{fig:delT} can be addressed well by the Schwarzschild precession of the orbit, which can modulate the recurrence timescale, especially for mildly eccentric orbits. Using the formula for the periapse (Schwarzschild) precession, the timescale for the apsidal rotation by $90^{\circ}$ can be expressed as follows,
\begin{equation}
 \tau_{\frac{\pi}{2}}\sim 80.3 \left(\frac{P_{\rm orb}}{2.2\,\text{days}} \right)^{\frac{5}{3}} \left(\frac{M_{\bullet}}{10^{5.8}\,M_{\odot}} \right)^{-\frac{2}{3}} \left(\frac{1-e^2}{0.96} \right)\,\text{days}\,, 
\end{equation}
where $e$ is the orbital eccentricity, which is set to $e=0.2$. This implies that the orbital orientation with respect to the accretion disk changes on the timescale of 100 days due to the periapse precession and this can thus partially address the changing eruption recurrence. For instance, initially, for a highly inclined perturbing body, the orbital orientation when the perturber intersects the disk at the apoapse and the periapse has the eruption periodicity of $\sim 1.1$ days or half of the orbital period. In $\tau_{\frac{\pi}{2}}\sim 80$ days, both intersections are close to the periapse, specifically the eruption recurrence timescale is $0.82$ days for $e=0.2$. {\bf This is comparable to the change in the recurrence timescale between Swift 1 and 2 datasets, see Figure~\ref{fig:delT}.} Since the Swift monitoring was separated by a few months and lasted always only a few days, for each monitoring session we effectively capture the system in a specific orbit-disk orientation and hence the recurrence timescale would vary accordingly. 

\begin{figure}
    \centering
    \includegraphics[width=\columnwidth]{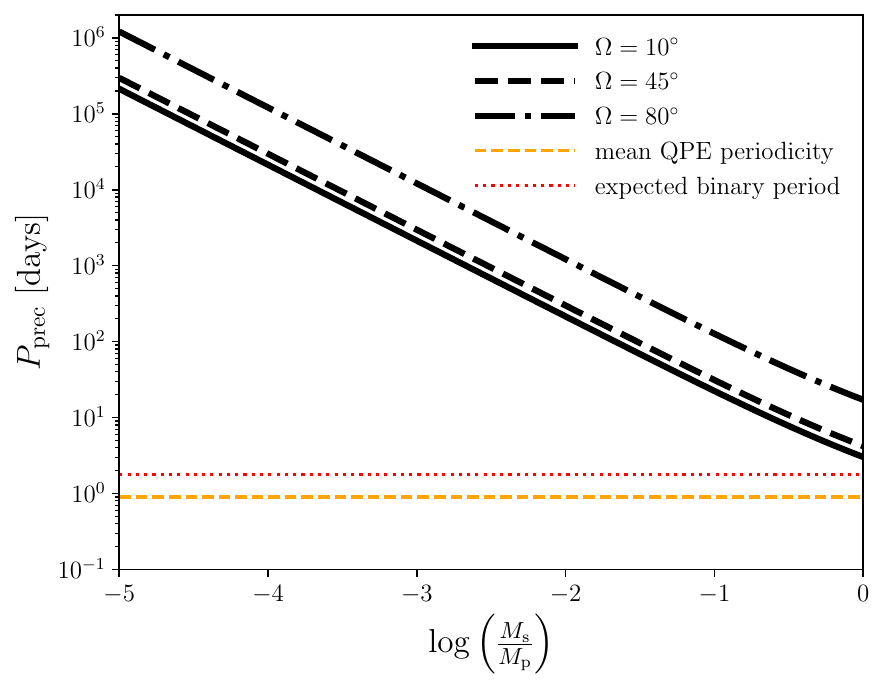}
    \caption{{\bf Accretion disk's precession period due to the torques by the secondary (expressed in days) as a function of the logarithm of the component mass ratio ($\log{(M_{\rm s}/M_{\rm p})}$). We depict three cases corresponding to the low inclination of the secondary with respect to the accretion disk ($\Omega=10$ degrees, solid line), intermediate inclination ($\Omega=45$ degrees, dashed line), and high inclination of $\Omega=80$ degrees (dot-dashed line). The mean QPE periodicity and the expected binary period are represented by dashed orange and red dotted lines, respectively.}}
    \label{fig_precession_ratio}
\end{figure}

Further variations can be caused by the disk precession \citep[see also][]{alessiamodel}, which can take place on various timescales depending on the outer radius of the misaligned disk and the black hole spin when the disk precession is rigid-like and driven by the Lense-Thirring effect. {\bf The Lense-Thirring precession also leads to the rotation of the line of nodes for the orbiting body that also affects the timing between the eruptions, especially for perturbers close to the rotating SMBH.} 

The disk precession can also be driven by the torques from the misaligned object and the precession period in that case is longer than the orbital period. {\bf The disk precession due to the torques from the orbiting secondary has a timescale of a few hundred days to days for massive perturbers of at least $\sim 10^4\,M_{\odot}$ for the total mass (primary$+$secondary) of $10^6\,M_{\odot}$. More specifically, for the secondary to primary mass ratio of $M_{\rm s}/M_{\rm p}=10^{-2}$, the precession period is $214$ days and for $M_{\rm s}/M_{\rm p}=10^{-1}$ the precession period is 22 days, assuming the inclination of $\Omega=10$ degrees between the secondary and the accretion disk around the primary. In Fig.~\ref{fig_precession_ratio}, we plot the expected precession period as a function of the mass ratio adopting the model of \citet{1995MNRAS.274..987P}, \citet{1997MNRAS.290..490L}, and \citet{2018MNRAS.478.3199B}. We consider three inclinations (10, 45, and 80 degrees), a polytropic index for the non-relativistic gas ($n=3/2$), the semi-major axis of $\sim 0.014$ milliparsecs (corresponding to the orbital period of $\sim 1.8$ days), and the outer radius of the precessing disk corresponding approximately to the semi-major axis of the binary. We see that the precession period of a few days is only possible for nearly equal-mass components.      }

Another reason {\bf for changes in periodicity} could be the time lag between the perturber-disk interaction and the time when the perturbation reaches the SMBH. The thermal front propagation is given by the sound-crossing timescale as $t_{\rm front}\approx R/(\alpha c_{\rm s})\sim (H/R)^{-1} t_{\rm th}$, where the thermal timescale can be estimated as $t_{\rm th}\sim \alpha^{-1} [R^3/(GM_{\bullet})]^{1/2}\sim 0.36 (\alpha/0.1)^{-1} (R/100 R_{\rm g})^{3/2} (M_{\bullet}/10^{5.8})\,\text{days}$. For the scale-height to radius ratio of $H/R\sim 0.1$, the front propagation timescale is  $t_{\rm front}\sim 3.6 [(H/R)/0.1]^{-1}(\alpha/0.1)^{-1} (R/100 R_{\rm g})^{3/2} (M_{\bullet}/10^{5.8})\,\text{days}$. Therefore the eruption recurrence timescale can be modulated by the propagation timescale that is further affected by the disk thickness, and hence the current accretion state. 

While in the perturber-disk interaction model presented by \citet{petra} the eruptions are produced by quasi-periodic enhancements in the accretion rate, in the analogous models presented by \citet{itaimodel2} and \citet{alessiamodel}, the X-ray emission flare is produced in shocked, optically thick expanding clouds of disk material ejected above and below the disk \citep[e.g., see discussion in][]{itaimodel2}. In principle, both emission mechanisms -- accretion-based and shock-based -- could be at work. The decreasing amplitude of eruptions in Fig. \ref{Fig:fig3} is consistent with models of \citet{petra, alessiamodel, itaimodel2}. It can be attributed to  diminishing inclination between the perturber's orbit and the accretion disk, which decreases the {\bf relative velocity of the perturber with respect to the disk material, and hence} the energy generated in density waves and shocks. Such an alignment process can take place due to the ongoing Bardeen-Petterson effect if the accretion disk is initially misaligned with respect to the equatorial plane. {\bf Alternatively, it could also be the result of the disk surface density becoming lower due to an ongoing decrease in the accretion rate. This follows from the eruption luminosity being proportional to the disk surface density, $L_{\rm QPE}\propto \Sigma$ \citep{itaimodel2,2023MNRAS.526...69T}, and for the standard disk there is a power-law dependency of the surface density on the accretion rate, $\Sigma\propto \dot{m}^{7/10}$ \citep{2002apa..book.....F}. The accretion rate can be decreasing over the course of several months to years following a TDE \citep[see e.g.][and further discussion]{itaimodel2} with a power-law time-dependency, in particular $\dot{m}\propto t^{-5/3}$ for the canonical TDE, and hence $L_{\rm QPE}\propto t^{-7/6}$. } One of the key differences between models proposed by \citet{petra} and those by \citet{alessiamodel} and \citet{itaimodel2} is that the former predicts the presence of repeated outflows. However, the search for such outflows is beyond the scope of this letter. 

As discussed in \cite{itaimodel2}, the emission properties of QPEs may secularly evolve due to changes in the accretion flow, and specifically the disk's scale-height and accretion rate, as well as changes to the secondary object's physical radius. If the secondary is a star, its outer layers are repeatedly ablated by shocks at every disk passage, consequently changing the star's cross section and its interaction with the disk. \cite{itaimodel2} further proposed that the origin of the accretion disk is bound debris of a previously tidally disrupted star around the same SMBH. The long term evolution of the TDE disk naturally results in evolution timescales of order years-decade (e.g., their equations 34-35), in agreement with the trends observed in the data. Another outcome of star-disk interaction is the orbital decay induced by hydrodynamical drag at disk passages. {\bf However, this should lead to a rather slow, gradual decrease in the recurrence timescale unlike a rather abrupt change of the recurrence timescale from 1.1 days in June, 2021 to 0.8 days in October, 2021, i.e. over the course of four months. The stellar orbit can only be abruptly changed by a massive mass loss from the system, for example, when a distorted stellar body would split like in a Hills mechanism. }

{\bf The star-disk interaction model was also presented by \citet{2023MNRAS.526...69T}, where the QPE luminosity is dominated by the breakout emission of the bow-shock of the star as it emerges from the AGN disk. They argue that for eRO-QPE1 both the breakout emission as well as the cooling emission of the expanding shocked bubble could contribute. This could partially address multiple peaks in Lomb-Scargle periodograms since the breakout emission should contribute once per orbit (when we see the emerging bow shock as the star ascends above the accretion disk), while we detect the cooling emission twice due to the expansion of the shocked gas both above and below the accretion disk. On the other hand, their model requires massive stars ($\sim 10\,M_{\odot}$) on retrograde, low-inclination orbits, which appears to be rather restrictive.  }

Finally, it is important to note that the shape of the eccentric orbits with a smaller pericenter distance near a fast-spinning black hole can be very different from Keplerian-like ellipses \citep[e.g.][chapt.~7]{1998mtbh.book.....C}. In fact, the disk-crossing radius can fluctuate in a rather wide range, thus giving a possibility to explain changes of the outburst properties \citep{1994ApJ...422..208K,xianmodel,2023MNRAS.526L..31K}.

\subsubsection{QPE from Roche-lobe overflow from a star orbiting an SMBH}
Roche-lobe overflow from an orbiting star was suggested to explain several QPE properties, such as their peak luminosities, temperatures, and the flare durations \citep{krolikmodel,itaimodel1} . In that case a small fraction of the stellar envelope is tidally stripped close to the pericenter of the stellar orbit. Therefore the eruption recurrence timescale is given by the orbital period, which sets the semi-major axis to $a\sim 251.6 (P_{\rm orb}/0.9\,\text{days})^{2/3}(M_{\bullet}/10^{5.8}\,M_{\odot} )^{-2/3} R_{\rm g}$. The X-ray eruptions could be produced via the {\bf oblique} stream-stream shocks close to the innermost stable circular orbit, which can address the soft X-ray thermal emission {\bf corresponding to $\sim 110$ eV}. The stellar orbit is expected to be only mildly eccentric ($e\lesssim 0.5$). In that case, for the soft X-ray eruptions to be produced close to the innermost stable circular orbit, the detached stellar stream needs to lose {\bf its} specific angular momentum that is initially comparable to the orbital angular momentum of the star at the pericenter crossing. This can be achieved by several processes, such as induced magnetic stresses \citep{krolikmodel} and/or magnetohydrodynamic drag due to ambient hot plasma \citep{2014A&A...565A..17Z}, especially  when the stellar orbit is more compact on the scale of a few $10\,r_{\rm g}$, i.e. the model can thus work better for eRO-QPE1 when the SMBH is heavier with $\sim 10^{6.4}\,M_{\odot}$. Assuming that most of the energy is dissipated at $R_{\rm dis}\sim 10\,R_{\rm g}$ due to {\bf oblique} stream-streamshocks driven by apsidal precession, the mean stellar mass loss per eruption can be estimated as $\Delta M\sim 7\times 10^{-7}\,(L_{\rm erupt}/2.2\times 10^{42}\,{\rm erg\,s^{-1}})(\tau_{\rm erupt}/8\,\text{h})(R_{\rm dis}/10\,R_{\rm g})\,M_{\odot}$, where $L_{\rm erupt}$ is the mean eruption luminosity and $\tau_{\rm erupt}$ is the mean eruption duration. The system SMBH--star is relatively long-lived since the stellar body is depleted in $\tau_{\star}=(M_{\star}/\Delta M)P_{\rm orb}\sim 3500$ years. The merger timescale for the orbital period of $P_{\rm orb}\sim 0.9$ days and a nearly circular orbit is $\tau_{\rm merge}\sim 4.85 (P_{\rm orb}/0.9\,\text{days})^{8/3}(M_{\bullet}/10^{5.8}\,M_{\odot})^{-2/3}(M_{\star}/1\,M_{\odot})^{-1}$ Myr, hence about three orders of magnitude longer than $\tau_{\star}$. The irregularity of the eruption recurrence {\bf can be caused by the intense X-ray irradiation of the upper stellar atmosphere. This stems from the comparison of the X-ray and stellar flux densities at the periapse \citep{krolikmodel}, 
\begin{align}
F_{\rm X}/F_{\star}&\sim 6\times 10^3 (1-e)^{-2} (L_{\rm erupt}/2.2 \times 10^{42}\,{\rm erg\,s^{-1}}) \times\,\notag\\
&(P_{\rm orb}/0.9\,\text{days})^{-4/3}
(M_{\bullet}/10^{5.8}\,M_{\odot})^{-2/3} (T_{\star}/T_{\odot
})^4\,.
\label{eq_ratio_xray_star}
\end{align}
The maximum unbound mass per orbit escaping the star due to the excess heating by the X-ray eruption can be estimated as follows,
\begin{align}
\Delta M_{\rm unbound}&\sim 3\times 10^{-8} (1-e)^{-2} (L_{\rm erupt}/2.2\times 10^{42}\,{\rm erg\,s^{-1}})\times \,\notag\\
&(D_{\rm erupt}/0.3) (P_{\rm orb}/0.9\,\text{days})^{-1/3} \times \,\notag\\
&(M_{\bullet}/10^{5.8}\,M_{\odot})^{-2/3} (M_{\star}/1\,M_{\odot})^{1.64}\,M_{\odot}\,,
\label{eq_unbound_mass}
\end{align}
where $D_{\rm erupt}$ is the duty cycle of eruptions. For mildly eccentric orbits ($e\sim 0.5$), $\Delta M_{\rm unbound}$ is comparable to $\Delta M$, and hence the modulation of mass transfer and QPE luminosity via the excess heating is a plausible mechanism.
}
The stellar mass transfer may thus be initially enhanced and unstable {\bf due to the induced turbulence in the upper atmosphere}. {\bf In addition, the progressive tidal truncation of the star may lead to its shrinking when the atmosphere layer is depleted via the Roche-lobe overflow.} The short periodicity of $\sim 0.9$ days is consistent with a Solar-like star whose radius is comparable to the Roche-lobe radius at its pericenter. The required stellar radius has an approximately Solar value. {\bf Using the relation for the limiting stellar radius for the tidal stripping, $R_{\star}\sim G^{1/3}/(4\pi^2)^{1/3}(1-e) P_{\rm orb}^{2/3}M_{\star}^{1/3}$, we obtain the numerical estimate of  $R_{\star}/R_{\odot}\sim 3.92 (1-e)(P_{\rm orb}/0.9\,{\rm days})^{2/3} (M_{\star}/1\,M_{\odot})^{1/3}$. Hence, if the stellar radius slightly shrinks below this value, the flares may be weakened or fall below the detection limit. The stripping of the envelope can also be connected with the inspiral of the star, and thus shortening of the QPE period \citep{2014ApJ...788...99B}. However, the precise timing of these processes requires 3D hydrodynamical models involving a realistic stellar model that is subject to tidal forces and radiation heating due to the X-ray flare close to the pericenter. } \\

{\bf A modification of the Roche-lobe overflow model was presented by \citet{wenbinmodel}, where the Roche-lobe overflowing star feeds a compact accretion disk. As in \citet{krolikmodel}, the QPE emission is due to circularization shocks, hence not accretion dominated. The steady-state accretion disk should provide a quiescent accretion with the Eddington ratio of $\sim 0.08$ (their Eq. 12) for $M_{\bullet}=10^6\,M_{\odot}$, which is above the inferred upper limit on the quiescent X-ray luminosity of eRO-QPE1.}

\subsection{QPE models related to accretion flow instabilities}
\subsubsection{QPEs from radiation pressure instability}
Accretion disk instabilities were invoked already in early studies \citep{Lightman1974} and revealed that the innermost regions of viscous disks cannot be stable when reaching a significant fraction of Eddington luminosity.  In the unstable mode, the radiation pressure becomes so strong that the local cooling exceeds the heating. When the local temperature increases, the excess of accretion rate leads to the disk depletion and density drops. The disk can rebuild on short timescales and enter a cyclic oscillatory mode, if the advective process is regulating the thermal imbalance. Numerical simulations of such oscillations presented in \cite{Janiuk2002} and the theoretical lightcurves well matched the observations of Galactic microquasars, GRS 1915+105, and then IGR J17091 (see \citealt{Janiuk2015}).
The extension of the unstable zone, and hence amplitudes and timescales of the observable luminosity flares, depend on the mass inflow rate, and also on the mass of the central black hole.
Application of the same global disk instability model to a wide range of black hole masses, from stellar mass to intermediate and SMBH, is quite straightforward. However, in order to produce realistic patterns, certain modifications of disk physics have to be assumed. For instance, \cite{Mikolaj2017} considered alternative forms of the viscous stress tensor, and found out that oscillation periods on the order of 1 day are possible for an IMBH of a 4$\times$10$^{4}$ M$_{\odot}$. The ratio of flare width to the recurrence time between flares is another measurable quantity, that may help to verify this model and its parameters.
Alternatively, \cite{marzenamodel} discussed explicitly the role of magnetic fields in regulating the oscillation pattern and partial stabilisation of the disk. For the case of $10^{5} M_{\odot}$ black hole, the intra-day oscillation timescales, characteristic of the QPE phenomenon, are possible with a strong magnetic field.
For the accretion rate of $\dot m$ = 0.5, and 
magnetic field of coefficient $b=0.22$ decreased the outburst timescale to $\sim 
$days. In this case, the pattern is irregular, and few-day flares are accompanied by sequences of one-day outbursts.
In addition, changing the viscosity parameter to $\alpha=0.1$ shortened the outburst timescale down to $16$ hours.
We note that the shape of the outbursts obtained from radiation pressure instabilities is asymmetric and the dimming phase takes 10\% of the duration of the full flare, which appears to be in contradiction with fast rise-slow decay or symmetric profiles of most of the flares of \target. {\bf In addition, considering the upper limit on the quiescent-level flux of \target, $F_{\rm q}\lesssim 4 \times 10^{-14}\,{\rm erg\,s^{-1}\,cm^{2}}$, the Eddington ratio can be constrained to be $\dot{m}\lesssim (4 \pi D_{\rm L}^2 F_{\rm q})\kappa_{\rm bol}/L_{\rm Edd}\sim 0.03$ for $M_{\bullet}=10^{5.8}\,M_{\odot}$ and the bolometric correction of $\kappa_{\rm bol}\sim 10$ \citep{2006ApJS..166..470R,2019MNRAS.488.5185N}. For the \xmm's quiescent flux of $F_{\rm q}^{\rm XMM}\sim 5\times 10^{-15}\,{\rm erg\,s^{-1}\,cm^{-2}}$, we obtain $\dot{m}\sim 0.004$. Hence, the relative accretion rate is at least one order of magnitude lower than the limiting accretion rate of $\dot{m}_{\rm RPI}\gtrsim 0.16 \alpha_{0.1}^{41/29} m_{5.8}^{-1/29}$ that is required for the radiation-pressure instability to effectively operate \citep{2020A&A...641A.167S} within the radiation-pressure dominated standard accretion disk.}

\subsubsection{QPEs from shock front oscillations}
Another possible source of QPEs is  oscillations of the shock front in low angular momentum flows. The source of the low angular momentum material accreting onto the SMBH can be strong stellar winds from massive stars orbiting the central black hole on larger distances ($\sim$ parsec scale) as was estimated for Sgr A* by \citet{2006MNRAS.370..219M}. If the accreting material on the central SMBH has sub-Keplerian angular momentum distribution, for a certain range of the parameter space (energy and angular momentum of the incoming matter) the possibility of multiple critical points existence appears \citep{1981AcA....31..283P,1982AcA....32....1M,1989ApJ...336..304A}.
In such a case, a shock front connected with a sudden drop of the inward velocity from supersonic to subsonic regime and simultaneous increase of the flow density may emerge. The accretion solution then passes through both the outer and the inner sonic points. The relation of quasi-spherical slowly rotating accretion flows to variability of X-ray sources has been reported already by \cite{1988ChA&A..12..119J} and the parameter space corresponding to the shock emergence in different geometrical setup was described by \cite{1990ApJ...350..281A}. 

More recently, numerical simulations have shown, that for a subset of parameters, the shock front location is unstable and oscillations of the shock bubble develop \citep{2015MNRAS.447.1565S}, which is accompanied by quasi-periodic flares in the accretion rate. \citet{2017MNRAS.472.4327S} provided extended study of the dependence of the oscillations on angular momentum and energy by means of 1D/2D and 3D GRMHD simulations, while \citet{2019MNRAS.487..755P} focused on the effect of adiabatic index on the resulting flow. 

Considering the mass of the SMBH in the range $M \in (10^{5.2},10^{6.4}) M_{\odot}$, the recurrence time of 0.9 days or $(10^5 - 6 \times 10^3)\,GM/c^3$ in geometrized units corresponds to the frequency $f \in (1 \times 10^{-5}, 2 \times 10^{-4})\,c^3/(GM)$. Such values were reported by \citet[see Table 3 and Fig. 18]{2017MNRAS.472.4327S} for shock fronts oscillating at the distance of several tens of gravitational radii from the center.

The model can accommodate the variations  in the period taking into account that the oscillations shown by \citet{2017MNRAS.472.4327S} are quasi-periodic in nature even when the parameters of the incoming gas (energy and angular momentum) are kept strictly constant. Moreover, the mean position of the shock front and the frequency of the oscillations depend quite strongly on those parameters, hence their relatively small change can lead to a change in the recurrence period or even to the disappearance/reappearance of the shock front, which is reflected in the accretion rate behaviour. The gradual decrease of the peak flux may be attributed to the slow density decline of the incoming gas. 

The coherent oscillations of the shock front are expected to appear in a situation where the incoming gas has relatively stable properties falling inside the multicritical parameter space, which may be a short-lived situation depending on the properties of the gas on the larger scale.

\subsubsection{QPEs from disk tearing instability}
The disk tearing instability can occur in disks that are sufficiently misaligned to a spinning black hole such that a large warp amplitude can develop in the disk \citep{Nixon:2012,Dogan:2018,Raj:2021a}. When only a modest warp amplitude is reached, the warp propagates through the disk either via waves if the disk is hot and low viscosity or diffusion if the disk is cool and high viscosity \citep{Papaloizou:1983}. However, for disks that achieve a large warp amplitude it has been shown that they can become unstable, leading to the disk separating into diskrete rings \citep[][see also \citealt{Ogilvie:2000}]{Dogan:2018}. In either the wavelike or diffusive case of warp propagation it is possible for the disk to tear into discrete rings \citep{Drewes:2021}. In disks that are strongly unstable, the disk emission can exhibit complex variability as the unstable region can be far from the black hole; this means that a combination of shocks, subsequent accretion and geometric effects are responsible for determining the emergent lightcurve. These processes are discussed in detail in \cite{Nixon:2014} and \cite{Raj:2021b}. 

However, for disks that are only weakly unstable, the instability is confined to the very inner regions of the disk. \cite{Raj:2021b} provide an example simulation of such a disk, where the innermost annulus of the disk is repeatedly torn off and accreted in a brief flare-like event. In their Fig.~3 they provide the accretion rate on to the central black hole with time, which shows regularly spaced peaks. They suggest that X-rays may be produced when the ring of gas that is torn from the disk precesses and shocks against its neighboring ring before falling into the black hole. For a black hole of mass $4\times 10^6\,M_\odot$ the spacing between these peaks is approximately a day, and thus could be consistent with the eruptions in \target. The shape of most of the eruptions observed in \target (being fast rise and slower decay) are reversed compared to those seen in Fig.~3 of \cite{Raj:2021b}; while this does not appear consistent, it may be that a more detailed model of the emission from the disk could account for this shape. For example, the shocks may cool slower than assumed in the simulation or they may be optically thick such that some expansion is required before the radiation may escape.

It is also possible that the X-ray flux declines over time in a tearing disk, as observed for \target (Fig. \ref{Fig:fig3}), if either the disk accretion rate or inclination are declining with time so that there is less energy generated in the shocks; \textbf{if the disk was formed from a past TDE, for example, the monotonically declining fallback rate (as $\propto t^{-5/3}$ for complete disruptions) would naturally lead to a weakening X-ray flux with time owing to the diminishing mass supply}. This changes the disk conditions which, in turn, leads to a change in the radius at which the disk tears and hence a change to the recurrence timescale. The most significant change in the time between eruptions for \target occurred between Swift \#1 and Swift \#2, and this drop in recurrence timescale was accompanied by the largest drop in average peak flux. This could be explained by a (small) reduction in the radius at which the disk tears (perhaps caused by the disk becoming slightly thinner) such that less material is involved in the inter-ring shocks. It is difficult to see why the period would then increase again while the flux continues to decline. However, to within the error bars the period is consistent with being constant from Swift \#2 to \#6, and therefore this might be reasonably explained by the radius of the instability remaining roughly constant while the disk accretion rate drops slightly, reducing the average peak flux accordingly. To confirm whether such details can be adequately reproduced by a disk tearing model requires targeted simulations. Such simulations could be used in the future to constrain the disk and black hole parameters for \target in the case that disk tearing is driving the eruptions. 

\textbf{
\subsection{Summary of models}
\begin{enumerate}
    \item{\textbf{The repeated partial stripping of a white dwarf (\citealt{zalamea10, kingqpemodel}) requires such a highly relativistic pericenter that, in spite of the extremely small mass ratio, gravitational-wave emission non-trivially reduces the orbital period over the timescale of 3 years. The non-detection of gravitational-wave decay, as our data implies (see Figure \ref{fig:lsps}), is only consistent with this model if the white dwarf mass is substantially below the (observationally constrained) most likely mass of $0.6 M_{\odot}$, and/or if the black hole mass is $\lesssim 10^4 M_{\odot}$ and in the intermediate-mass black hole regime}.}
    \item {\bf The interaction of the stellar or the compact-object perturber with a standard accretion disk \citep{itaimodel2,alessiamodel} can address the eruption luminosity, its temperature, flare recurrence timescale, and its irregularities (due to the various precession mechanisms). The decrease in the amplitude seems to require a previous TDE, which ensures that the accretion rate is progressively getting smaller.}
    \item {\bf The Roche-lobe overflow from the main-sequence star does not necessarily require the existence of the standard accretion disk (\citeauthor{krolikmodel}, \citeyear{krolikmodel}; see, however, \citeauthor{wenbinmodel}, \citeyear{wenbinmodel}), hence the model is also suitable for low-luminosity sources such as eRO-QPE1. It can also explain the flare luminosity, recurrence timescale, and temperature when the detached stream can collide with itself close to the innermost stable circular orbit, which results in oblique circularization shocks \citep{krolikmodel}. The caveats include the angular momentum loss for the matter to reach the innermost stable circular orbit. The recurrence irregularity can be addressed by the additional mass loss due to X-ray heating. The decrease in the flare amplitude and the period could be related to a rapid mass loss from the disturbed stellar body though this would require a detailed numerical modelling.}
    \item {\bf Models involving disk instabilities depend on the accretion rate, viscosity parameter of the accretion flow, and its magnetic field strength and  configuration. For the radiation-pressure instability to operate effectively, the relative accretion rate should at least reach $\dot{m}\sim 0.1$ \citep{2020A&A...641A.167S} so that the inner part of the disk is radiation-pressure dominated and thus unstable. Therefore, this model appears problematic for eRO-QPE1, whose upper limit on the Eddington ratio is $\sim 0.01$. Another model involving the oscillating shock bubble modulating the accretion rate in a quasiperiodic manner \citep{2015MNRAS.447.1565S,2017MNRAS.472.4327S} depends on the boundary conditions (e.g. a distant wind-blowing star) supplying low angular-momentum material to the inner regions. The model requires the fine-tuning of several parameters to ensure that the period of the QPE as well as the eruption amplitude decrease as observed. Finally, the disk-tearing instability can address the timing and the quasiperiodic manner of the eRO-QPE1 flares by the mechanism of the detachment and the precession of the inner disc ring and its collision with the neighbouring ring, resulting in bright flares due to shocks and subsequent accretion \citep{Raj:2021b}. The disc-tearing instability requires a sufficiently high misalignment between the accretion disc and the SMBH spin to operate. A targeted simulation is necessary to address the change in the QPE period as well as the amplitude with time.  }
\end{enumerate}
}
\section{discussion}
\label{sec:discussion}

We note that in accreting  Stellar-mass Black Holes (StMBHs), Quasi-Periodic Oscillations (QPOs) of the X-ray flux have been known for several decades \citep{mcclin,2006csxs.book.....L}. There are intriguing similarities as well as differences compared to QPOs in low-mass X-ray binaries, nonetheless, in the case of accreting neutron stars, properties of QPOs are likely determined by the internal oscillations within the accretion flow and in the boundary layer. In the case of StMBH QPOs, timescales are in the range of a fraction of a second to a few milliseconds \citep{1998AdSpR..22..925V,mcclin}. Thus, a typical power spectrum consisting of a few ks of exposure samples several tens to hundreds of thousands of cycles of the underlying phenomenon. In the case of eRO-QPE1, using roughly 300~ks of XRT exposure we sampled about 27 eruptions. Thus, it is possible that we are looking at individual frequencies/timescales making up the broad quasi-periodicity similar to those seen in StMBHs, and it may not be valid to make strong inferences based on time between individual eruptions.

\subsection{Comparison to GSN~069 and repeating TDEs}
\label{sec:gsn}
There are two aspects of eRO-QPE1 that are strikingly similar to GSN~069's behavior. First, the recurrence time varies between 0.6~d to 1.2~d (Fig.~\ref{fig:lsps}). This corresponds to a coherence value, Q, defined as the ratio of dispersion in time between eruptions and the mean duration between eruptions, of 0.9~d/0.3~d $\approx$3. This is comparable to GSN~069 where the recurrence time varied between 25 to 35 ks before they disappeared in 2020 (see the bottom right panel of Fig.~3 of \citealt{miniutti23a}), i.e., coherence of 30/10 $\approx$3. 

Secondly, GSN~069 showed a decline in QPE intensity over a period of 500 days following the first detection of eruptions (see Fig.~2 of \citealt{miniutti23a}). \target is showing the same trend (Fig.~\ref{Fig:fig3}). More recent follow-up of GSN~069 has shown that QPEs have disappeared for about 2 years before reappearing with a much shorter recurrence period \citep{alivenkicking}. Continued monitoring of \target will test if \target continues to behave the same way.

Two X-ray TDEs, AT2018fyk/ASASSN-18UL \citep{wevers23} and eRASSt J045650.3-203750 \citep{liu23} are known to repeat on timescales of $\sim$1200 and $\sim$200 days, respectively. These have been interpreted as rpTDEs \citep{wevers23, liu23}. It is worth noting that these two systems show a diminishing luminosity in successive peaks, indicating a progressive reduction in peak amplitude over time, but we note that only two peaks were observed for AT2018fyk and 4 for eRASSt~J045650.3-203750 (obtained in private communication).

\subsection{Future Prospects of Tracking eRO-QPE1's Eruptions:}\label{sec:future}
Based on the mean duration of eRO-QPE1 eruptions of 7.6 hours \citep{arcodia21} and the capabilities of current X-ray facilities, i.e., \xmm, \nicer, and \swift we estimate a sensitivity limit beyond which detecting eRO-QPE1's eruptions would be challenging. For instance, \swift has an orbital period of roughly 96 mins (5.6 ks) around the Earth and can typically observe a target for a few ks per orbit (e.g., see \url{https://swift.gsfc.nasa.gov/proposals/cy20\_faq.html\#monitor}). Assuming 2 ks exposure every 5.6~ks, on average, \swift can accumulate about 10~ks over a 7.6 hour duration of a typical eruption. In order to detect an eruption robustly and roughly constrain its temperature, one would need at least 25 counts, i.e., a 5$\sigma$ threshold. This translates to an average count rate during an eruption of 25/10~ks = 0.0025 count s$^{-1}$. Using HEASARC WebPIMMS (\url{https://heasarc.gsfc.nasa.gov/cgi-bin/Tools/w3pimms/w3pimms.pl}) and assuming a blackbody temperature of 0.11 keV, this translates to an observed 0.3-1.2 keV flux of 6$\times$10$^{-14}$ \cgs. In other words, if the average eruption flux is below this value, \swift/XRT would find it difficult to clearly identify an eruption. Similarly  for \nicer, assuming a minimum count rate of 2x the nominal threshold value of 0.2 count s$^{-1}$ \citep{3c50}, implies that if the average eruption flux is below 3$\times$10$^{-14}$ \cgs, \nicer will find it  difficult to identify the eruptions. Thus, detecting eruptions with \nicer should already be challenging in the present state (see the bottom right panel of Fig.~\ref{fig:lc} and top panel of Fig.~\ref{Fig:fig3}). For \xmm, the prospects are better. For example, if we require to detect at least 50 counts within 7.6 hours, the corresponding count rate is $\sim$0.002 count s$^{-1}$, which translates to a flux of 2$\times$10$^{-15}$ \cgs. Thus, a 1.2 d or 125 ks of continuous exposure with \xmm can guarantee detecting an eruption. We do not consider \chandra due to its deteriorating soft X-ray response. In summary, if the decaying trend in eruption peak luminosity shown in Fig.~\ref{Fig:fig3} continues, they can be traced for the next several years with both \swift and \xmm.

\section{Summary}\label{sec:summary}

\target is the second QPE source showing a gradual decay in strength of eruptions over time. In the case of GSN~069, \citet{alivenkicking} have recently reported that the QPEs decayed and eventually disappeared for about two years and re-appeared with a much shorter recurrence time. Also, in the case of GSN~069, along with the eruption strength, the quiescent level was also declining over time. The \swift/XRT observations presented here were not sensitive enough to detect the quiescent level. Only \xmm has the effective area and the sensitivity to detect and track the quiescent level over the coming years. It is unknown if \target's quiescence and eruptions  exhibit the same behavior. Further monitoring observations with \swift and \xmm over the next few years will certainly be able to address this question and may enable a unifying picture for long-term evolution of QPE sources. 

\begin{acknowledgments}
D.R.P was funded by a NASA grant 80NSSC22K0090 to conduct this study. ERC acknowledges support from the National Science Foundation through grant AST-2006684. MZ acknowledges the financial support of the GA\v{C}R Junior Star grant GM24-10599M. VK thanks to the Czech Science Foundation (ref.\ 21-06825X). PS has been supported by the fellowship Lumina Quaeruntur No.\ LQ100032102 of the Czech Academy of Sciences. This work was supported by the Ministry of Education, Youth and Sports of the Czech Republic through the e-INFRA CZ (ID:90140). CJN acknowledges support from the Science and Technology Facilities Council (grant number ST/Y000544/1) and the Leverhulme Trust (grant number RPG-2021-380).
MK acknowledges support from DLR grant FKZ 50 OR 2307.
MS acknowledges support from Polish Funding Agency National Science Centre, project 2021/41/N/ST9/02280 (PRELUDIUM 20), the European Research Council (ERC) under the European Union’s Horizon 2020 research and innovation program (grant agreement 950533) and from the
Israel Science Foundation, (grant 1849/19). IL acknowledges support from a Rothschild Fellowship and The Gruber Foundation.
AJ was supported by grant No. 2019/35/B/ST9/04000 from the Polish National Science Center, Poland, and also wants to acknowledge support from the PL-Grid infrastructure and Interdiskiplinary Center for Mathematical Modeling of the Warsaw University.
\end{acknowledgments}
\vspace{5mm}
\facilities{\swift (XRT), \xmm/EPIC}

\appendix
\setcounter{table}{0}
\renewcommand{\thetable}{A\arabic{table}}
\setcounter{figure}{0}
\renewcommand{\thefigure}{A\arabic{figure}}

\section*{Supplementary material}
\section{Data Reduction and Analysis}\label{sec:data}
We used data from \swift/XRT and \xmm's European Photon Counting Camera (EPIC) pn in this work. The details of our data reduction procedures are discussed below.

\subsection{\swift/XRT}\label{sec:xrt}
Between June 2021 and June 2023 \swift's X-Ray Telescope (XRT; \cite{swift, xrt}) performed six sets of high-cadence observations of \target. These were part of \swift's approved guest observer programs: 1720147 (cycle 17), 1821153 (cycle 18) and 1922142 (cycle 19) (PI: Pasham). In each of these six campaigns, \target was observed for 12-16 times per day for 4-5 days. Each visit was between 100-500 s long with a cumulative exposure of about 50~ks per campaign.

We started our analysis by downloading the raw data from the publicly available HEASARC archive: \url{https://heasarc.gsfc.nasa.gov/cgi-bin/W3Browse/w3browse.pl}.  Using the {\tt xrtpipeline} tool from HEASoftv6.29c we reduced these 353 datasets as per the standard data reduction guidelines as described on \swift webpages: \url{https://www.swift.ac.uk/analysis/xrt/xrtpipeline.php}. A few of the obsIDs had exposures of just a few tens of seconds and those were excluded from further analysis. From each of the remaining obsIDs, we extracted source counts from a circular region with a radius of 30$^{\prime\prime}$ centered on coordinates: (02:31:47.26, -10:20:10.31) (J2000.0 epoch). Background events were extracted from an annulus centered on the same coordinates with an inner and outer radius of 70$^{\prime\prime}$ and 220$^{\prime\prime}$, respectively. We only used events with grades 0-12. We combined data on per-obsID basis and extracted the resulting background-subtracted count rates (Figure \ref{fig:lc}).

\textbf{To track the strength of the eruptions over the past 3 years we employ two different methods. First, for each \swift data set (\swift\#1...6), we assign a data point as belonging to an eruption if the count rate is more than 10 times the mean background rate which is roughly 0.0003 cps in all cases. Then we compute the mean of the top 20\% of all count rates assigned to eruptions. This value is a proxy for the peak X-ray flux. These estimates are shown as filled circles in Fig. \ref{Fig:fig3}. }

\textbf{Independently, we also compute the average X-ray flux of eruptions per \swift dataset as follows. First, we use the Bayesian blocks algorithm of \citealt{blocks2} to identify the start and the end times of each eruption in each dataset. For this, we use the {\tt Astropy} \citealt{2013A&A...558A..33A} implementation of this algorithm (\url{https://docs.astropy.org/en/stable/api/astropy.stats.bayesian\_blocks.html}) which also takes the measurement errors into account. As evident by eye, the eruptions are especially weak in the later epochs. Thus, following the description in section 4.3 of \citealt{blocks2}, we use a lenient false alarm probability of 50\% to improve sensitivity to identify eruptions. The corresponding bin edges are shown as black dashes in Fig. \ref{fig:lc}. Then, we estimate the average strength of the eruptions as the mean of all count rates belonging to eruptions in a given epoch.  The resulting curve is shown as filled triangles in Fig. \ref{Fig:fig3} and is consistent with analysis based on peak rates derived from a simple count rate cut described above. }

\textbf{The Bayesian blocks algorithm also allows us to identify the peaks time of each eruption. For 
 \swift\#1, these values are  0.66$\pm$0.03 d, 1.82$\pm$0.07 d, and 2.92$\pm$0.10 d for the 3 eruptions identified by the algorithm. The times are measured with reference to the first observation, i.e., MJD 59373.074. For \swift\#2, the peak times of eruptions (with reference to MJD 59502.022) are 0.34$\pm$0.10 d, 1.17$\pm$0.14 d, 1.93$\pm$0.03 d, 2.73$\pm$0.10 d, 3.56$\pm$0.13 d. For \swift\#3, these values are 0.11$\pm$0.11 d, 0.72$\pm$0.09 d, 1.54$\pm$0.07 d, and 3.24$\pm$0.17 d. The algorithm did not identify any eruptions during \swift\#4. During \swift\#5, the eruption peaks occurred at 0.76$\pm$0.13 d, 1.34$\pm$0.10 d, 3.20$\pm$0.03 d, and 3.93$\pm$0.02 d. It is evident from these values that the recurrence time between eruptions can vary between 0.6 to 1.2 d. This is consistent with the locations of the broad peaks (0.6-1.2 d) in the Lomb Scargle periodograms shown in Fig. \ref{fig:lsps}.
}

To convert from observed 0.3-1.2 keV count rate to flux/luminosity we extracted a combined X-ray spectrum using data from all eruptions in each campaign, modeled it with a thermal component ({\it tbabs*ztbabs*zashift(diskbb)} and obtained a conversion factor (see Table \ref{tab:tab1}). Similarly, we constrained the quiescent flux level by combining obsIDs between eruptions in each campaign. The 3-$\sigma$ count rate/flux upper limits were obtained using {\tt ximage}'s {\tt sosta} tool.

\subsection{\xmm/EPIC}\label{sec:epic}
\xmm's European Photon Imaging Camera (EPIC) observed \target on two occasions, 27 July 2020 (obsID: 0861910201) and 4 Aug 2020 (0861910301) for about 90~ks each. We downloaded the raw data from the HEASARC archive and reduced it using the standard procedures: \url{https://www.cosmos.esa.int/web/xmm-newton/sas-threads}. We extracted source events from a circular aperture of 25$^{\prime\prime}$ centered on coordinates mentioned above and background events were extracted from a nearby  circular region of 50$^{\prime\prime}$ radius. We ensured that the background aperture was free from point sources. Flaring windows were removed by filtering on the 10-12 keV light curve of the entire field of view. ObsID 0861910201 had two clear eruptions while one eruption was present in obsID 0861910301. 

We extracted the EPIC-pn X-ray spectra of the three peaks of the eruptions and binned them using \xmm analysis tool {\tt specgroup} to have a minimum of 1 count per spectral bin with {\it oversample=3}. Then we fit them separately with the same model used for modeling \swift/XRT eruptions, i.e., {\it tbabs*ztbabs*zashift(diskbb)}. While the temperature and the normalization of {\it diskbb} were left free, the neutral column density of the host was tied across all 3 spectra. The fit resulted in C-stat/dof of 83.3/65. The best-fit column density was (0.069$\pm$0.020)$\times$10$^{22}$ cm$^{-2}$. The best-fit temperature values (in chronological order) were 0.125$\pm$0.006 keV, 0.192$\pm$0.010 keV, and 0.121$\pm$0.005 keV. The corresponding 0.3-1.2 keV observed fluxes were (8.4$\pm$0.3)$\times$10$^{-13}$ \cgs, (33.8$\pm$0.8)$\times$10$^{-13}$ \cgs, and (7.6$\pm$0.3)$\times$10$^{-13}$ \cgs, respectively. Because these observations were closely spaced in time, and we are interested in the evolution over years timescale, we report the average values in Fig.~\ref{Fig:fig3}. \textbf{Similarly, we computed the average flux of the eruptions. }

We then extracted two quiescence spectra, one from each of the obsIDs and fit them again with {\it tbabs*ztbabs*zashift(diskbb)}. We fixed the host galaxy neutral column density at 0.069$\times$10$^{22}$ cm$^{-2}$ and tied the {\it diskbb}'s temperature and normalization which resulted in a C-stat/dof of 35.5/33. The best-fit temperature was 0.082$\pm$0.025 keV. The observed 0.3-1.2 keV flux could not be constrained independently. However, if we freeze the disk temperature at the best-fit value of 0.082 keV, the resulting flux is (4.7$\pm$0.7)$\times$10$^{-15}$ \cgs.


\bibliography{ms}{}
\bibliographystyle{aasjournal}

\end{document}